\documentclass[onecolumn, draftcls, 12pt]{IEEEtran}
\IEEEoverridecommandlockouts
\usepackage{cite}
\ifCLASSINFOpdf
  \usepackage[pdftex]{graphicx}
  \DeclareGraphicsExtensions{.pdf,.jpeg,.png}
\else
\usepackage[dvips]{graphicx}
\DeclareGraphicsExtensions{.eps}
\fi
\usepackage[cmex10]{amsmath}
\usepackage{amssymb}
\usepackage{graphicx}
\usepackage{amsthm}
\usepackage{algorithmic}
\usepackage{array}
\usepackage{mdwmath}
\usepackage{mdwtab}
\usepackage{balance}
\usepackage{color}
\usepackage{multirow}
\usepackage[tight,footnotesize]{subfigure}
\usepackage{url}
\graphicspath{{./Images/}}
\allowdisplaybreaks
\begin{document}
\title{Harvesting Resource Allocation in Energy Harvesting Wireless Sensor Networks}
\author{Shenqiu~Zhang,~\IEEEmembership{Student~Member,~IEEE,}~and~Alireza~Seyedi,~\IEEEmembership{Senior~Member,~IEEE}%
\thanks{S. Zhang is with the ECE Dept., University of Rochester, Rochester, NY, e-mail: shenqiu.zhang@rochester.edu, A. Seyedi is with the EECS Dept., University of Central Florida, Orlando, FL, email: alireza.seyedi@ieee.org.}
\thanks{This work has been funded in part by the National Science Foundation (NSF) grant CNS-1049691.}
}
\maketitle

\vspace{-0.75in}
\begin{abstract}
Considering an energy harvesting sensor network, the overall probability of event loss is derived. Based on this result, a variety of harvesting resource allocation schemes (sizing the energy storages and the harvesting devices, under a total cost constraint) are provided. Their performances are verified and compared through simulations.
\end{abstract}

\section{Introduction}\label{Chap4Sec1_Introduction}

Despite many advances in energy efficient communication techniques for wireless sensor networks (WSN),  the limitation of energy supply is still a critical issue. One of the most attractive solutions to this problem is energy harvesting \cite{Kansal2007}, where in each node a harvesting device harvests energy from ambient sources such as light, wind or vibration, and stores it in an energy storage device such as a rechargeable battery or a supercapacitor. Energy harvesting networks pose many new challenges due to the fact that their energy supply is dynamic and stochastic. 

A good body of work is being developed on harvesting-aware communication techniques and protocols that take the energy variability into account \cite{Murthy2009,Kar2006,Medepally2010,Seyedi2010,Sharma2010,Ozel2011,Devillers2012,Michelusi2012}. What has been less studied, is the design of energy harvesting nodes and network. In particular, harvesting resources, namely the sizes of the harvesting device and the energy storage, can have a significant impact on the performance of the network and, thus, must be carefully chosen. 

An approach to the design of energy storage capacity is developed in \cite{Reddy2010}, where it is assumed that in a particular time slot, the harvesting power and the consumed power are constant and known. In practice, however, available harvesting energy and the energy consumption are unpredictable and random in nature. In \cite{Kimball2009}, the authors take an empirical approach in which the capacity of energy storage and the capability of harvesting device are chosen based on the historical record of harvested power and predefined power consumption patterns. In \cite{Zhang13}, we have provided an analytical approach to the sizing of harvesting and storage devices, assuming that the energy and traffic processes are Markovian. 

All these works, however, consider the design of a single node and do not take the network aspects into account. In \cite{Zhang2011}, we considered a network with linear topology and provided optimal and sub-optimal solutions for the allocation of harvesting resources. Here, we extend our previous work to energy-harvesting sensor networks with an arbitrary topology. First, we derive the overall probability of loss for event reports. Then, we use this result to develop harvesting resource allocation schemes among the nodes, constrained by total cost. 

\section{System Model}\label{Chap4Sec2_Model}
Consider a wireless sensor network consisting of a total of $V$ nodes. The first $V-1$ nodes are energy-harvesting sensors and the $V$th node is the sink, which has access to unlimited energy. The sensors send their event reports, according to predetermined routing paths towards the sink. The (fractional) routing is represented by a weighted and directed graph with weighted adjacency matrix $\mathbf{R}=[r_{ji}]$, where $r_{ji}$ is the fraction of traffic in node $i$ routed to node $j$. Clearly, if there are no links from $i$ to $j$, we have $r_{ji}=0$. Furthermore, $\mathbf{R}$ is a left  stochastic matrix, that is $\sum_{j=1}^Vr_{ji}=1$. We will also define the binary adjacency matrix, $\mathbf{A}=[a_{ji}]$, where $a_{ji}=0$ if $r_{ji}=0$, and $a_{ji}=1$ otherwise. Note that $r_{iV}=a_{iV}=0$, since the sink has no outgoing links.

Node $v$, generates event reports at the rate $\lambda_v$, with $\lambda_V=0$ for the sink. The event reports are routed through the network and are forwarded to the sink. That is, each node is a source and may be a relay for traffic from upstream nodes. Event reports may be lost due to two factors: (i) shortage of remaining energy at a node (with probability $p_v$), and (ii) channel impairments of each link (with probability $q$). 

\section{Analysis}\label{sec:analysis}
\subsection{Traffic Flow and Total Event Loss}
Denote the rate of event reports arriving at node $j$ from node $i$, including reports generated at node $i$ as well as relayed reports originated at other nodes, by $\theta_{ji}$. Of course, no transmission exists if no link connects node $i$ to node $j$. Thus, the rate from $i$ to $j$ can be more clearly represented by $a_{ji} \theta_{ji}$. Therefore, the total rate of event reports arriving at node $v$ is
\begin{equation}\label{chap4eq1}
\theta_v = \lambda_v + \sum_{i=1}^{V} a_{vi} \theta_{vi}.
\end{equation}
Considering losses, the total outgoing rate from node $v$ is
\begin{equation}\label{chap4eq2}
\theta_{v}(1-p_v)(1-q) = \sum_{j=1}^{V} a_{jv} \theta_{jv},
\end{equation}
where the channel losses are absorbed into the originating node. We also have
\begin{equation}\label{chap4eq3}
\theta_{jv} = r_{jv} \theta_{v} (1-p_v) (1-q).
\end{equation}
Substituting \eqref{chap4eq3} into \eqref{chap4eq1} and \eqref{chap4eq2} yields
\begin{equation}\label{chap4eq4}
\theta_v = \lambda_v + (1-q)\sum_{i=1}^{V} r_{vi} \theta_{i}(1-p_i).
\end{equation}

For a matrix representation, we denote the generation rate of event reports by $\boldsymbol{\lambda} = [\begin{matrix} \lambda_{1} & \cdots & \lambda_{V}\end{matrix}]^T$, the rate of event reports arriving at nodes by $\boldsymbol{\theta} = [\begin{matrix} \theta_{1} & \cdots & \theta_{V} \end{matrix}]^T$, and the loss probabilities at sensor nodes due to the shortage of energy by matrix $\mathbf{P} = \textrm{diag}([p_1, \cdots, p_{V-1}, 0])$. Then, the matrix form of \eqref{chap4eq4} is $\boldsymbol{\theta} = \boldsymbol{\lambda} + (1-q) \mathbf{R} \left( \mathbf{I}- \mathbf{P}\right) \boldsymbol{\theta}$. Solving for $\boldsymbol{\theta}$ yields 
\begin{equation}\label{chap4eq9}
\boldsymbol{\theta} = \left[ \mathbf{I} - \left(1-q \right) \mathbf{R} \left( \mathbf{I} - \mathbf{P} \right)\right]^{-1} \boldsymbol{\lambda}.
\end{equation}

Now, $\theta_V=\mathbf{e}_V^T\boldsymbol{\theta}$ gives the rate of event reports arriving the sink, where $\{\mathbf{e}_v\}$, is the standard basis. Also, the total generation rate of event reports is given by $\mathbf{1}^T \boldsymbol{\lambda}$. Thus, the probability of losing event reports before reaching the sink is
\begin{eqnarray}\label{chap4eq11}
P_L = 1 - \frac{\theta_{V}}{\mathbf{1}^T \boldsymbol{\lambda}}= 1 - \frac{\mathbf{e}_V^T \left[ \mathbf{I} - \left(1-q \right) \mathbf{R} \left( \mathbf{I} - \mathbf{P} \right)\right]^{-1} \boldsymbol{\lambda}}{\mathbf{1}^T \boldsymbol{\lambda}}.
\end{eqnarray}

\subsection{Event Loss Due to Energy Shortage}
The above analysis provided the relationship between event rates and the overall event loss. One key factor, not discussed so far, is the relationship between loss probability at an energy-harvesting sensor node $p_v$, its total event rate $\theta_v$, and its capability of harvesting energy from the environment. To model the node loss probability $p_v$, we use the concept of ``energy packet''. That is, we assume that an agent fills the energy (from the continuous time process) into packets, and releases them once they are full. With this model, the energy arrival process becomes a point process, which is described by the inter-arrival times of the energy packets. Then, we can view the energy storage as a queue holding harvested energy packets and supplying them to the transmitter when necessary. We assume that each event report requires one energy packet for processing and transmission. Since incoming events are lost at a node when its energy queue is empty, the probability of an empty energy storage is exactly the same as the probability of losing events due to energy shortage. 

Assume that the energy packet process at node $v$ is Poisson with rate $\mu_v$. Also assume the event report arrival to be a Poisson process, which implies that the energy consumption process of node $v$ is also a Poisson process with rate $\theta_v$.  The energy queue in node $v$ can then be viewed as a $M|M|1|N_{v}$ queue  \cite{Kleinrock1975}, where $N_v$ is the capacity of energy storage of node $v$ in energy packets. Thus,
\begin{equation} \label{chap4eq13}
p_v = \frac{1 - \frac{\mu_v}{\theta_v}}{1-\left(\frac{\mu_v}{\theta_v}\right)^{N_v+1}}.
\end{equation}

\section{Harvesting Resource Allocation}\label{Chap4Sec3_Design}
The sizes of harvesting device and energy storage are represented by  the harvesting rate, $\mu_v$, and energy storage capacity, $N_v$, respectively. Assume that the total available amounts of harvesting rate and storage capacity, limited by total cost, are $\mu (V-1)$ and $N (V-1)$, respectively. Distribution of these harvesting resources among nodes has a significant impact on the network loss probability. In this section we study different strategies for allocation of these harvesting resources.

\subsection{Uniform Resource Allocation}
The simplest approach is to allocate the resources uniformly among all nodes. That is, $\mu_v^\dag=\mu$ and $N_v^\dag=N$. The probability of event loss is then given by \eqref{chap4eq9} and \eqref{chap4eq11} where
\begin{eqnarray}
p_v &=& \frac{1 - \frac{\mu}{\theta_v}}{1-\left(\frac{\mu}{\theta_v}\right)^{N+1}}.
\end{eqnarray}
Although this approach is simple, it does not perform well, due to the bottlenecks formed at the nodes closer to the sink which have considerably higher traffic, but the same resources.

\subsection{Optimal Resource Allocation}\label{Chap4Sec3_Optimal}
Ideally, one would allocate the resources such that $P_L$ is minimized. Due to \eqref{chap4eq11}, minimizing the network loss probability is equivalent to maximizing the event arrival rate at the sink $\theta_V$. Thus, an optimization problem can be formulated as
\begin{eqnarray}\label{chap4eq14}
\begin{array}{ll}
\underset{\mu_v,N_v}{\mbox{maximize}} & \qquad \mathbf{e}_V^T \left[ \mathbf{I} - \left( 1- q \right) \mathbf{R} \left( \mathbf{I} - \mathbf{P} \right) \right]^{-1} \boldsymbol{\lambda} \\
\textrm{subject to} & \qquad  \mathcal{C}_1: \,\,\, \frac{1}{V-1}\sum_{v=1}^{V-1} \mu_{v}= \mu\\
                    & \qquad  \mathcal{C}_2: \,\,\, \frac{1}{V-1}\sum_{v=1}^{V-1} N_{v} = N \nonumber \\
                    & \qquad  \mathcal{C}_3: \,\,\, p_v = \frac{1 - \frac{\mu_v}{\theta_v}}{1-\left(\frac{\mu_v}{\theta_v}\right)^{N_v+1}} \nonumber \\
                    & \qquad  \mathcal{C}_4: \,\,\, \theta_v = \mathbf{e}_v^T \left[ \mathbf{I} - \left( 1- q \right) \mathbf{R} \left( \mathbf{I} - \mathbf{P} \right) \right]^{-1} \boldsymbol{\lambda}.
\end{array}
\end{eqnarray}
The solutions of this optimization problem, $\mu_v^\star$ and $N_v^\star$, can be found using classic optimization algorithms such as adaptive simulated annealing. However, numerical approach is time-consuming and does not provide a useful design perspective. In the following, we propose a simple sub-optimal approach.

\subsection{Almost-Fair Resource Allocation}\label{Chap4Sec3_Fair}
The optimal resource allocation discussed above usually yields an unfair performance. That is, sensors close to the sink will have relatively smaller node loss probabilities than those far from the sink. To overcome this unfairness, we propose a simple resource allocation by requiring equal node loss probability for all nodes. Note that sensors may still have different probability of loss for their own generated data, which is why we call this scheme \emph{almost-fair}.

By enforcing $p_v = p$, \eqref{chap4eq9} becomes
\begin{eqnarray}\label{chap4eq15}
\boldsymbol{\theta} &=& \left[ \mathbf{I} - \left(1-q \right)\left(1-p\right) \mathbf{R} \right]^{-1} \boldsymbol{\lambda} \nonumber\\
&=& \sum_{n=0}^{V-1} \left(1-q \right)^n\left(1-p\right)^n \mathbf{R}^n \boldsymbol{\lambda},
\end{eqnarray}
where we have utilized the fact that $\mathbf{R}$ is nilpotent, i.e. $\mathbf{R}^n = \mathbf{0}$ for $n \ge V$. Thus, our problem reduces to finding $\mu_v$ and $N_v$ subject to $\mathcal{C}_1$, $\mathcal{C}_2$, and
\begin{eqnarray}\label{chap4eq16}
\begin{array}{ll}
\mathcal{C}'_3: \,\,\, p = \frac{1 - \frac{\mu_v}{\theta_v}}{1-\left(\frac{\mu_v}{\theta_v}\right)^{N_v+1}} \\
\mathcal{C}'_4: \,\,\, \theta_v = \sum_{n=0}^{V-1} \left(1-q \right)^n\left(1-p\right)^n \mathbf{e}_v^T \mathbf{R}^n \boldsymbol{\lambda}.
\end{array}
\end{eqnarray}

To further simplify the problem, we limit the optimization domain to $\frac{\mu_v}{\theta_v}=\frac{\mu_w}{\theta_w}$ and $N_v = N_w$. For such a solution to satisfy the constrains $\mathcal{C}_1$ and $\mathcal{C}_2$ in \eqref{chap4eq16} we must have
\begin{eqnarray}\label{chap4eq17}
\frac{\mu^\ddag_{v}}{\theta_v} = \frac{(V-1)\mu}{\displaystyle\sum\nolimits_{w=1}^{V-1}\theta_w} \qquad \textrm{and} \qquad
N^\ddag_{v} =N.
\end{eqnarray}
We now need to solve for $\mu^\ddag_{v}$ such that  $\mathcal{C}'_3$ and $\mathcal{C}'_4$ are also satisfied. Substituting \eqref{chap4eq17} in $\mathcal{C}'_3$ yields
\begin{eqnarray}\label{chap4eq18}
p = \frac{1-\alpha}{1-\alpha^{N+1}},
\end{eqnarray}
where
\begin{eqnarray}
\alpha = \frac{\mu^\ddag_{v}}{\theta_v} = \frac{(V-1)\mu}{\displaystyle\sum\nolimits_{i=1}^{V-1}\theta_i}.
\end{eqnarray}
Substituting \eqref{chap4eq18} and $\mathcal{C}'_4$ into $\mu^\ddag_v = \alpha \theta_v$ yields
\begin{equation}\label{chap4eq19}
\mu^\ddag_v = \alpha \sum_{n=0}^{V-1} \left(1-q \right)^n\left(\frac{\alpha - \alpha^{N+1}}{1-\alpha^{N+1}}\right)^n \mathbf{e}_v^T \mathbf{R}^n \boldsymbol{\lambda}.
\end{equation}
Furthermore, substituting \eqref{chap4eq19} into $\mathcal{C}_1$ yields
\begin{eqnarray}\label{chap4eq20}
f(\alpha) & = & \alpha \sum_{n=0}^{V-1} \left(1-q \right)^n\left(\frac{\alpha - \alpha^{N+1}}{1-\alpha^{N+1}}\right)^n \sum_{v=1}^{V-1} \mathbf{e}_v^T \mathbf{R}^n \boldsymbol{\lambda} - \mu \left(V-1\right)=0.
\end{eqnarray}
We now need to solve \eqref{chap4eq20} and obtain its root $\alpha^\ddag$ which in turn yields $\mu^\ddag_v$ from \eqref{chap4eq19}. It is easy to verify that $f(.)$ is monotonically increasing, and that $f(0) = -\mu(V-1) < 0$. Thus, $f(\alpha)=0$ has exactly one positive root, which can be found using a binary search if we have a point $\alpha_+$ such that $f(\alpha_+)>0$. To find $\alpha_+$, let us assume $\alpha_+ \ge 1$. Then,  and it is easy to utilize \eqref{chap4eq18} to show that $p \le 1/(N+1)$ and $(\alpha_+-\alpha_+^{N+1})/(1-\alpha_+^{N+1}) = 1-p \ge N/(N+1)$. Therefore,
\begin{eqnarray}\label{chap4eq21}
f(\alpha_+) & \ge & \alpha_+ \sum_{n=0}^{V-1} \left(1-q \right)^n\left(\frac{N}{N+1}\right)^n \sum_{v=1}^{V-1} \mathbf{e}_v^T \mathbf{R}^n \boldsymbol{\lambda} - \mu \left(V-1\right)\nonumber\\
& = & \alpha_+   \sum_{v=1}^{V-1} \mathbf{e}_v^T \left[\mathbf{I}-\frac{\left(1-q \right) N}{N+1}\mathbf{R}\right]^{-1}\boldsymbol{\lambda} - \mu \left(V-1\right).\nonumber
\end{eqnarray}
Thus,
\begin{eqnarray}\label{chap4eq22}
\alpha_+ =  \max\left\{ 1, \frac{\mu \left(V-1\right)}{\left[\begin{array}{cccc}1 & \cdots & 1 & 0\end{array}\right] \left(\mathbf{I}-\frac{\left(1-q \right) N}{N+1}\mathbf{R}\right)^{-1}\boldsymbol{\lambda}} \right\}\nonumber
\end{eqnarray}
ensures that $f(\alpha_+)>0$. Therefore, the solution $\alpha^\ddag$ is readily found using a binary search over  $[0, \alpha_+]$. We note that this approach significantly reduces the computational complexity from a $2(V-1)$ dimensional non-convex problem to a one dimensional binary search. Once $\alpha^\ddag$ is found, $\mu_v^\ddag$ and $N_v^\ddag$ are determined using \eqref{chap4eq17} and \eqref{chap4eq19}.

\section{Results}\label{Chap4Sec5_Experiments}

In this section, we verify our analytical results using simulations and compare the performance of the resource allocation schemes proposed in Section \ref{Chap4Sec3_Design}.

\subsection{Simulation Setup}\label{Chap4Sec5_SimulationSetup}
The network is formed by random deployment of a total of $V-1$ energy harvesting sensors according to a uniform distribution over a disk. The sink node is located at the center of the disk. The connectivity of the nodes is then determined based on proximity. That is, two sensor nodes are connected if their distance is less than a radius $R$. Disconnected networks are discarded. After the nodes are deployed, their routing paths to the sink are determined using Dijkstra's shortest path algorithm, with links costs proportional to the square of the distance between the nodes, consistent with the free space path loss model. We note that while the analysis and the proposed harvesting resource allocation schemes cover the more general case of fractional routing, here, for simplicity, we limit ourselves to deterministic routes.

The simulation parameters of a sensor node, i.e. $\lambda_v$, $\mu_v$ and $N_v$, are set considering a ZigBee mote, MICAz \cite{XbowMICAz}, powered by solar energy \cite{Ho2010a}. Assuming that each event report consists of 10 packets, each consisting of $132$ bytes, the active time period to report one event is $56.96$ ms. With active operation power of $83.1$ mW \cite{XbowMICAz}, the energy required to transmit one event report is approximately $E = 4.73$ mJ. We assume that a NESSCAP $2.7$ V, $3$ F \cite{NESSCAPUltracapacitor} supercapacitors is used for the energy storage, whose storage capacity is $3$ mWh. This means that a fully-charged energy storage holds $2283$ energy units. We assume a harvesting power of $1.1$ mW \cite{Ho2010a}. Thus, the average rate of harvesting one energy unit is $0.2326$ Hz. To ensure that the event report generation rates are comparable to the load, we assume a typical event report generation rate of $0.4652/V$ Hz, for each sensor.

\subsection{Verifying Theoretical $P_L$}\label{Chap4Sec5_SimulationVsTheory}
\begin{figure}[!t]
\centering
\includegraphics[width=0.3\textwidth]{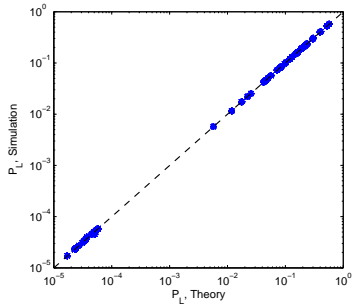}
\vspace{-0.25in}
\caption{Comparison between theoretical network probability of loss and simulation results.}\label{chap4fig2}
\vspace{-0.15in}
\end{figure}
This section compares the theoretical loss probability of a network $P_L$ with the simulation results. We randomly generate a large number of networks. For each network, the number of nodes is randomly taken from a uniform distribution over $10 \le V \le 100$. Moreover, each sensor node randomly selects its simulation parameters uniformly  within the $\pm 50\%$ range of the typical values given in Section \ref{Chap4Sec5_SimulationSetup}. The probability of loss due to channel impairment is assumed to be $q = 10^{-5}$. Fig. \ref{chap4fig2} compares the resulting network probability of loss to the analytical $P_L$ calculated by \eqref{chap4eq11} for $1482$ randomly generated sensor networks. We observe that the theoretical results matches the simulations quite well.

\subsection{Performance of the Harvesting Resource Allocation Schemes}\label{Chap4Sec5_CompareDesign}
This section compares the performance of the optimal, the almost-fair and the uniform harvesting resource allocation schemes. We randomly generate $1000$ networks, each of which contains $V=20$ nodes. We assume that all sensor nodes generate event reports at the same rate $\lambda_v = 0.4652/20 = 0.0233$ Hz.
\begin{figure*}[!t]
\centering
\begin{array}{cc}
\includegraphics[width=0.3\textwidth]{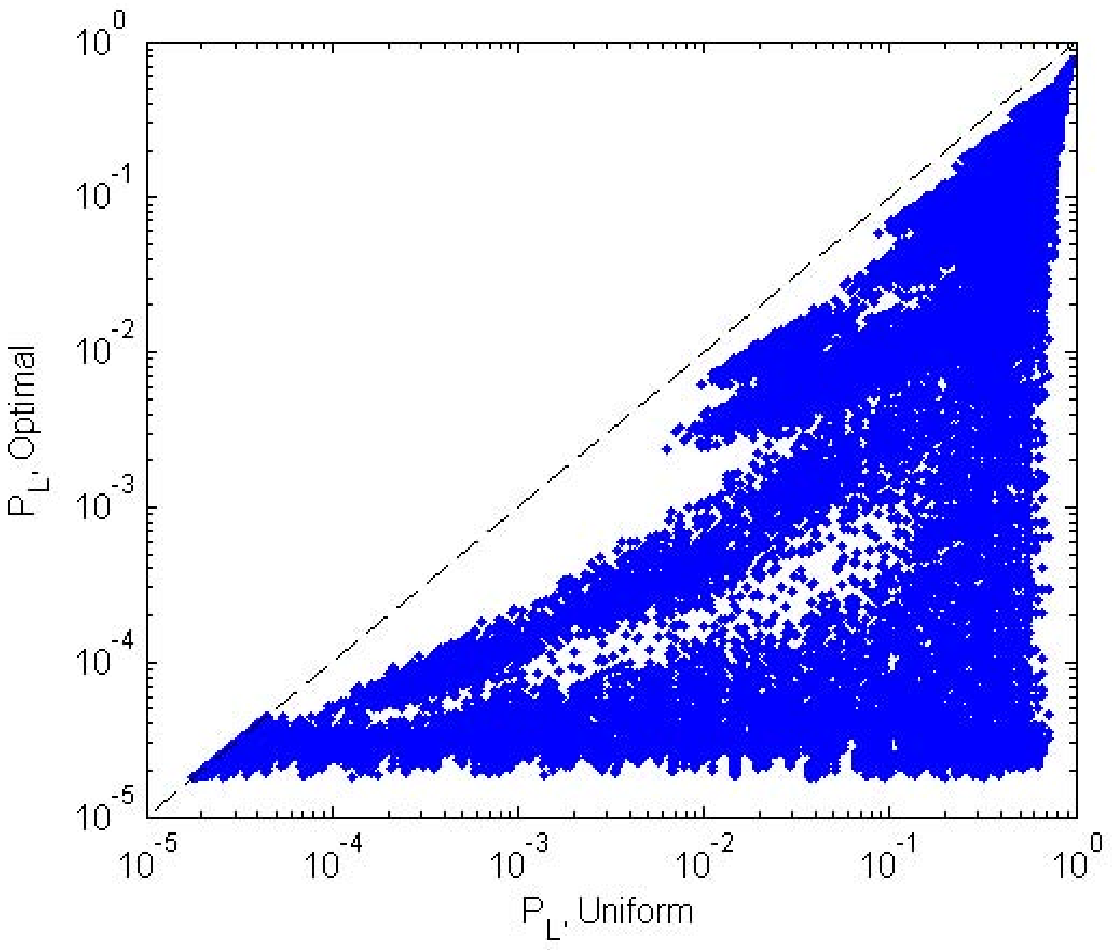}
&\includegraphics[width=0.3\textwidth]{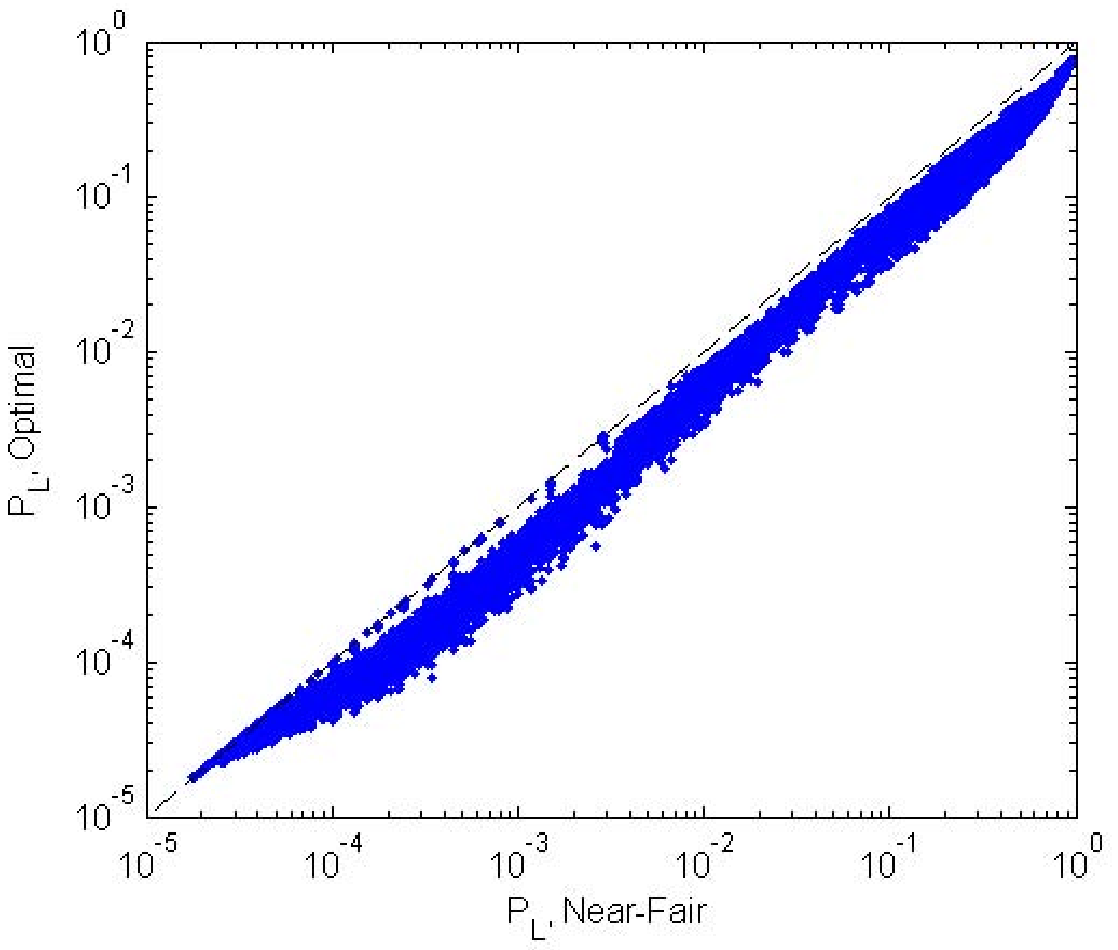}\\
(a)&(b)
\end{array}
\vspace{-0.1in}
\caption{Comparison of the network probability of loss for (a) optimal and uniform and (b) optimal and almost-fair resource allocation schemes}
\vspace{-0.35in}
\label{chap4fig3a}
\end{figure*}
For each network, we perform the three different allocation schemes with the same constrains. The constraints are drawn from a wide range of $1 \le N \le 10000$ and $0.01 \le \mu \le 10$. Fig. \ref{chap4fig3a}(a) compares the network probability loss for the optimal and uniform schemes. We can see that the uniform resource allocation performs considerably worse than the optimal. In average, the network probability of loss is higher by $2.2$ orders of magnitude. Fig. \ref{chap4fig3a}(b) provides a similar comparison between the optimal and the almost-fair schemes. In contrast to the uniform allocation, we see that the proposed almost-fair allocation performs well, and close to optimum. On average, the network probability of loss for the proposed almost-fair allocation is only worse than that of the optimal allocation by $0.15$ orders of magnitude.

\begin{figure*}[!t]
\centering
\begin{array}{cc}
\includegraphics[width=0.3\textwidth]{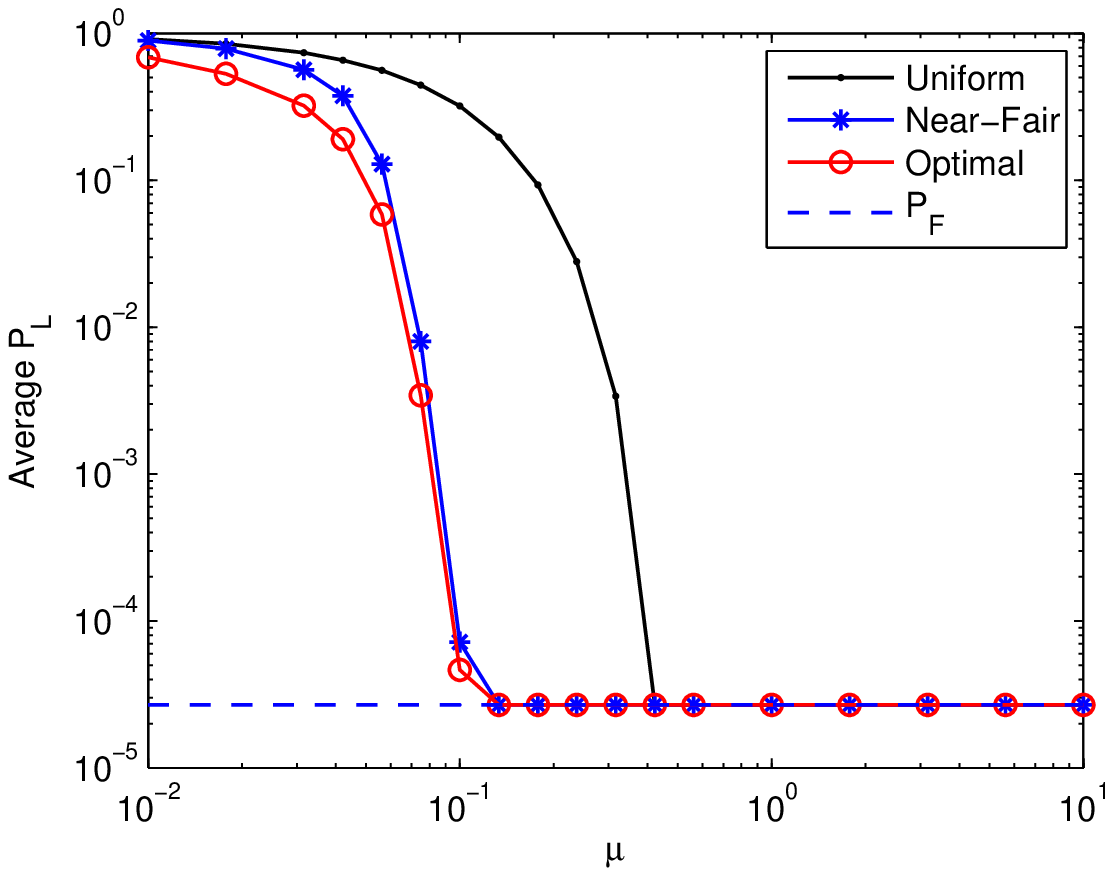} &
\includegraphics[width=0.3\textwidth]{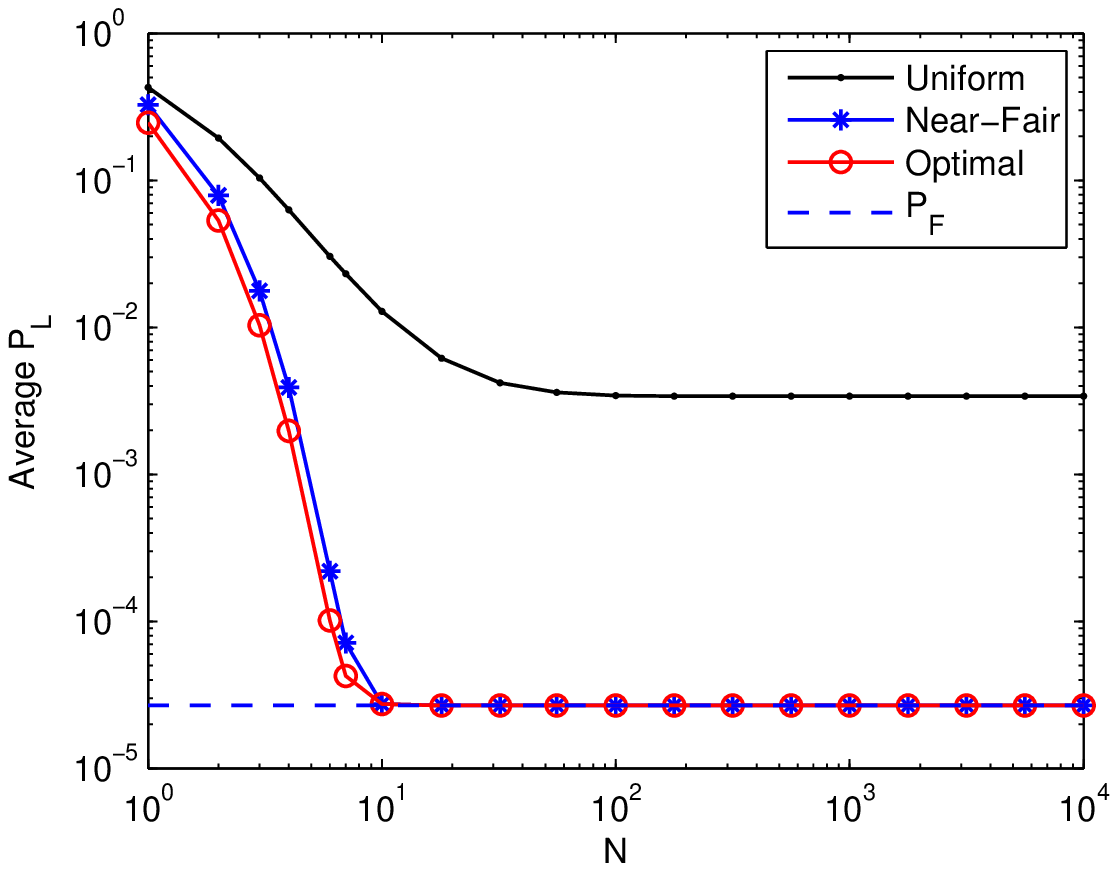}\\
(a) & (b) 
\end{array}
\vspace{-0.1in}
\caption{Network probability of loss (a) versus $\mu$ with $N=1000$ and (b) versus $N$ with $\mu=0.316$.}\label{chap4fig4}
\vspace{-0.35in}
\end{figure*}

For a different perspective, Fig. \ref{chap4fig4} depicts $P_L$ as a function of $\mu$ and $N$. From \ref{chap4fig4}(a) we observe that the uniform resource allocation requires higher levels of energy harvesting by a factor of approximately 4, compared to the optimal and almost-fair approaches. From Fig. \ref{chap4fig4}(b), we observe that the uniform approach never reaches the error floor for smaller $\mu$. On the other hand, the almost-fair approach has a very good performance compared with the optimal approach.

\section{Conclusions}\label{Chap4Sec6_Conclusions}
We have analyzes the loss probability of event reports in an energy harvesting WSN. Based on the analysis, an optimization problem for sizing of energy storages and harvesting devices is formulated. Moreover, we have proposed a simple almost-fair approach, which performs nearly as well as the optimal approach. Simulation results are utilized to verify the analytical results and compare the performances of the resource allocation schemes.

\bibliographystyle{IEEETran}
\bibliography{library}

\end{document}